\documentclass{aastex}
\usepackage{spr-astr-addons}
\usepackage{url}

\RequirePackage{color}

\begin{document}

\title{Relativistic anisotropic star and its maximum mass in higher dimensions}

\author{Bikash Chandra Paul}, \author{Pradip Kumar Chattopadhyay\altaffilmark{1}, Shibshankar Karmakar\altaffilmark{2}}
    
\affil{Physics Department, North Bengal University, \\
Siliguri, Dist. : Darjeeling, Pin : 734 013, West Bengal, India \\}
\and

\altaffiltext{1}{Physics Department, Alipurduar College,\\
P.O.: Alipurduar Court, Dist: Jalpaiguri, Pin: 736122, West Bengal, India}

\altaffiltext{2}{IUCAA Resource Centre, Physics Department, North Bengal University, India}

\begin{abstract}
We present a class of relativistic solutions of cold compact anisotropic stars in hydrostatic equilibrium in the framework of higher dimensions using spheroidal geometry. The solutions obtained with Vaidya-Tikekar metric are used to construct stellar models of compact objects and studied their physical  features.  The effects of anisotropy and extra dimensions on the global properties namely, compactness, mass, radius, equation of state  are determined in higher dimensions in terms of the spheroidicity parameter ($\lambda$). It is noted that for a given configuration, compactness of a star is found smaller in higher dimensions compared to that in four space-time dimensions. It is also noted that the maximum mass of  compact objects increase with the increase of space-time dimensions which however attains a maximum when  $D=5$ for a large ($\lambda=100$), thereafter it decreases as one increases number of extra dimensions. The effect of  extra dimensions on anisotropy is also studied.
\vspace{0.1 cm} \\

\end{abstract}

\section{Introduction}\label{Int}

During the last couple of decades  there has been  a considerable research activities in understanding
issues both in cosmology and in astrophysics in the framework of higher dimensions. Particularly the results obtained in the usual four dimensions are generalized in higher dimensions in addition to new physics. The history of higher dimensions goes back to the work done by 
 Kaluza and Klein in the past \citep{k1,k2}. Kaluza and Klein  independently first introduced the concept of extra dimension in addition to the usual four dimensions
to unify gravitational interaction  with that of electromagnetic interaction.  The theory is essentially
an extension of Einstein general theory of relativity (henceforth, GTR) in five dimensions
which is of much interest in particle physics as well as in  cosmology. But the initial approach
does not work well. A couple of decades ago the study of higher dimensional theories has been revived once again
and it was considerably generalized after realizing that many interesting theories of particle
interactions need more than four dimensions for their consistent formulation. On the other hand, GTR was formulated in a space-time
with just four  dimensions. Thus if some of the theories of particle interactions are consistent in higher dimensions, it is natural to
look for the generalization of the theories developed in the usual four dimensions. It
became important to generalize the results obtained in four dimensional GTR in the higher
dimensional context and probe the effects due to incorporation of one or more than one extra
space-time dimensions in the theory. In this direction Chodos and Detweiler \citep{cd} first obtained a  higher dimensional cosmological model and thereafter a number of cosmological models in higher dimensions have been discussed in the literature \citep{cd1,cd2,cd3,lp,cd6} to address different issues not understood in the usual four dimensions. In  cosmology it is proposed  that a higher
dimensional universe might undergo a spontaneous compactification leading to a product space
$M_4 \times M_d $, with $M_d$ the compact inner space,  describing the present universe satisfactorily.
Thereafter the advent of string theory \citep{str,str1,str2,str3} particularly a viable description of superstring
theory in 10 dimensions led to a spurt in activities in higher dimensions. The work
of Randall and Sundrum \citep{rs} led to a paradigm shift in understanding the compactification
mechanism. Randall and Sundrum gave an interesting picture of gravity in which the
extra dimensions is not compact and it is possible to recover the usual four dimensional Newtonian gravity from a five dimensional anti-de Sitter space-time in the low energy limit.
In the context of localized sources in astrophysics, higher dimensional versions of the spherically  symmetric Schwarzschild and Reissner-N$\ddot{o}$rdstrom black holes \citep{RN,RN1}, Kerr Black holes \citep{Kerr,Kerr1}, black holes in compactified space-time \citep{Myers}, no-hair theorem \citep{Sokolowski}, Hawking radiation \citep{Mperry}, Vaidya solution \citep{Iyer} have been generalized. Shen and Tan \citep{STan} also obtained a global regular solution of higher dimensional Schwarzschild space-time.  The mass to radius ratio in higher dimensions for a
uniform density star is determined which is a genaralization of the four dimensions and  new results have been reported in the literature \citep{bcp}. The
consequences of extra dimensions in understanding the structure of neutron stars employig Kaluza-Klein model
was investigated by Liddle {\it et al.} \citep{LiddleAR}. The model is constructed making use of a  five dimensional energy-momentum tensor
 described by perfect fluid. The four dimensional version of the theory is
found to have a perfect fluid with a scalar source. The effect of the source term is found very large
which  leads to a substantial lower value in the mass of neutron star associated with a particular
central density.

In astrophysics, it is known from recent observational  prediction that there exist a number of compact objects whose masses and radii are not compatible with the standard neutron star models. As densities of such compact objects are normally above the nuclear matter density, theoretical studies hints that pressure within such compact objects are likely to be anisotropic, {\it i.e.},  existence of two different kinds of interior pressures namely,  the radial pressure and the tangential pressure \citep{Herrera}. A number of  literature \citep{Thomas,Patel,Maartens,Gokhroo} came up where the solutions of Einstein's field equations with anisotropic fluid distribution on different space-time geometries are discussed. The role of pressure anisotropy is studied in the context of high-redshift values including the  stability of compact objects (see for example \citep{Mak,Dev,Chaisi} and references therein). Bowers and Liang \citep{Bowers} obtained the corresponding change in the limiting values of the maximum mass of compact stars in the presence of  anisotropy. Recently, the maximum mass  of an isotropic compact object and that of an  anisotropic one in the context of Vaidya-Tikekar model obtained by Karmakar {\em et al.} \citep{SK,SK1} in four dimension making use of general relativistic solution obtained by  Mukherjee {\em et al.} \citep{Mukherjee}.  It may be mentioned here that a higher dimensional generalization of the  relativistic solution obtained by  Mukherjee {\em et al.} \citep{Mukherjee} has been generalized by one of us (\citep{bcpaul}). In the present paper we estimate the maximum mass limit of an anisotropic compact object making use of the above general relativistic solution in higher dimensions.
In this case we consider  space-time geometry describe by  a metric ansatz given by Vaidya and Tikekar \citep{Vaidya}. The technique adopted here is different from that usually considered in obtaining relativistic solution from Einstein's field equation. Usually for a known equation of state (in short, EoS) of matter one obtains solution for the geometry. But in the case of compact objects the equation of state of matter inside a compact object is not yet known except some phenomenological assumptions. In this case making use of a  known geometry  for  compact objects in hydrostatic equilibrium in higher dimensional GTR we explore different physical features of the  compact objects. It helps to determine both the mass and radius of compact objects in terms of  geometrical parameters as was obtained in Ref. \citep{b1,b2} in four dimensions. We also predict the relevant  EoS  for a given configuration known from observations. The EoS obtained here satisfies a non linear equation. It may be mentioned here that similar non-linear EoS have been employed by  Mafa Takisa and  Maharaj \citep{maha} to obtain stellar models for compact objects.

The paper is organized as follows : In sec. \ref{Ani} we set up the Einstein field equation for an anisotropic star and presented  a class of  new solutions in higher dimensions. For physically relevant anisotropic stars, the regularity and matching conditions for the solution at the boundary of the star is ensured to obtain stellar models. In sec. \ref{Phy}  the role of anisotropy is studied and estimated the probable maximum mass for the class of solutions  obtained in section \ref{Max}. We conclude by summarizing our results in section \ref{Dis}.

\section{Field equation in Higher Dimensions and Solutions}\label{Ani}
The Einstein's field equation in higher dimensions is given by
\begin{equation}
{\bf R}_{AB}-{1\over2} g_{AB}{\bf R} = 8 \pi G_{D}T_{AB}\label{eq1}
\end{equation}
where D is the total number of dimensions, $G_{D}= G V_{D-4}$ is the gravitational constant in $D$ dimensions, $G$ denotes the $4$ dimensional gravitational constant and $V_{D-4}$ is the volume of extra space. ${\bf R}_{AB}$ is Ricci tensor, ${\bf R}$ is Ricci scalar, $g_{AB}$ is metric tensor and $T_{AB}$ is the energy momentum tensor in $D$ dimensions.
We consider the metric of a higher dimensional spherically symmetric, static space-time given by
\begin{equation}
ds^{2} = -e^{2\nu (r)}dt^{2}+e^{2\mu (r)}dr^{2}+r^{2}d\Omega^2_{n}\label{eq2}
\end{equation}
where $\nu(r)$ and $\mu(r)$ are the two unknown metric functions, $n=D-2$ and $d\Omega^2_{n}=d\theta^2_{1}+ sin^2\theta_{1}d\theta^2_{1}+sin^2\theta_{2}(d\theta^2_{3}+...+ sin^2\theta_{n-1}d\theta^2_{n})$ represents the metric on the $n$-sphere in polar coordinates. The energy-momentum tensor for an anisotropic star in the most general form is given by
\begin{equation}
T_{AB} = \mbox{diag}~(-\rho,~ p_{r},~p_{t},~p_{t},...,~p_{t})\label{eq3}
\end{equation}
where $\rho$ is the energy-density, $p_{r}$ is the radial pressure, $p_{t}$ is the tangential pressure and $\Delta = p_{t}-p_{r}$ is the measure of pressure anisotropy in this model, which depends on metric potential $\mu(r)$ and $\nu(r)$.
Using eqs. (\ref{eq2}) and (\ref{eq3}), Einstein's field equation reduces to the following set of equations:
\begin{equation}
8\pi G_{D}\rho = \frac{n(n-1)\left(1-e^{-2\mu}\right)}{2r^2}+\frac{n\mu'e^{-2\mu}}{r},\label{eq4}
\end{equation}
\begin{equation}
8\pi G_{D}p_{r} = \frac{n\nu'e^{-2\mu}}{r}-\frac{n(n-1)\left(1-e^{-2\mu}\right)}{2r^2},\label{eq5}
\end{equation}
\[
8\pi G_{D}p_{t}=e^{-2\mu} \left(\nu ^{\prime \prime }+{\nu ^{\prime }}^{2}-\nu ^{\prime
}\mu ^{\prime }-\frac{(n-1)(\mu ^{\prime }-\nu ^{\prime})}{r}\right)
\]
\begin{equation}
\; \; \; \; \; \; \; \; \; \; 
-\frac{(n-1)(n-2)\left(1-e^{-2\mu}\right)}{2r^2} \label{eq6}
\end{equation}
Using eqs. (\ref{eq5}) and (\ref{eq6}), pressure anisotropy condition ($\Delta = p_{t}-p_{r}$) gives rise to
\begin{equation}
\nu ^{\prime \prime }+{\nu ^{\prime }}^{2}-\nu ^{\prime}\mu ^{\prime }-\frac{(n-1)\mu ^{\prime }}{r}-\frac{\nu ^{\prime}}{r}-\frac{(n-1)\left(1-e^{2\mu}\right)}{r^2}=\Delta e^{2\mu}\label{eq7}
\end{equation}
To solve the eqs. (\ref{eq4})-(\ref{eq7}), we use the ansatz \cite{Vaidya},
\begin{equation}
e^{2\mu }=\frac{1+\lambda r^2/R^2}{1-r^2/R^2},\label{eq8}
\end{equation}
where $\lambda$ being the  spheroidicity parameter and $R$ is the geometrical parameter. Now from eq. (\ref{eq7}), one obtains a second order differential equation in $x$, given by
\[
(1+\lambda-\lambda x^2)\Psi_{xx} + \lambda x \Psi_{x} + \lambda(\lambda + 1)(n-1)\Psi
\]
\begin{equation}
\; \; \; \; \; \; \; 
 - \frac{\Delta R^2(1+\lambda - \lambda x^2)^2}{(1-x^2)} \Psi = 0\label{eq9}
\end{equation}
where $\Psi = e^{\nu(r)}$, with  $ x^{2} = 1-\frac{r^2}{R^2}$.\\
Now for simplicity we choose the anisotropic parameter $\Delta$ \cite{SK} as follows,
$$\Delta = \frac{\alpha \lambda^2(1-x^2)}{R^2(1+\lambda-\lambda x^2)^2}$$
The above relation is chosen so that the regularity at the centre of the star is ensured. The method adopted here to obtain solution of the field eqs. (\ref{eq4})-(\ref{eq7})  is similar to that previously obtained by Mukherjee {\it et. al.} \cite{Mukherjee}. Using the transformation $z = \sqrt{\lambda/(\lambda +1)} x$, eq. (\ref{eq9}) can be written as
\begin{equation}
(1-z^2)\Psi_{zz} + z\Psi_{z} + (\beta^2 - 1)\Psi  = 0\label{eq10}
\end{equation}
where $\beta =\sqrt{(n-1)(\lambda+1)-\lambda \alpha +1}$ is a constant. The general solution of eq. (\ref{eq10}) \cite{Mukherjee} is given by
\begin{equation}
e^{\nu }=A\bigg[{\frac{\cos [(\beta+1)\zeta +\delta]}{\beta+1}}-{\frac{ \cos [(\beta-1)\zeta+\delta]}{\beta-1}}\bigg] \label{eq11}
\end{equation}
where $\zeta =\cos ^{-1}z$. $A$ and $\delta$ are two constants which can be determined from the boundary conditions. For a real $\beta$ the anisotropy parameter $\alpha$ satisfies a limit determined by the space-time dimensions $(D)$ and spheroidicity parameter $\lambda$ which is  $\alpha_{max}<(D-3)+\frac{D-2}{\lambda}$. The physical parameters relevant in this model are given below:
\begin{eqnarray}
\rho &=& {n \over 16 \pi G_{D} R^2 (1-z^2)} \bigg[n-1 + {2 \over (\lambda + 1) (1
- z^2)} \bigg] \label{eq12}\\
p_{r} &=& - {1 \over 8 \pi G_{D} R^2 (1-z^2)} \bigg[ {n(n-1)\over 2} + {n z \Psi_{z}\over
(\lambda + 1) \Psi} \bigg] \label{eq13}\\
p_{t} &=& p_{r} + \Delta \label{eq14}\\
\Delta &=& {\alpha \lambda \over 8 \pi G_{D} R^2} \bigg[ {(\lambda + 1)(1 - z^2) - 1 \over (\lambda + 1)^2(1 - z^2)^2} \bigg] \label{eq15}
\end{eqnarray}
Eqs. (\ref{eq12}) - (\ref{eq15}) together with eqs. (\ref{eq8}) and (\ref{eq11}) will be employed here to obtain  exact solution of the Einstein field equation. The mass of a compact star of radius $b$ \cite{bcpaul} in higher dimensions is given by
\begin{equation}
M(b) =\frac{nA_{n}}{16\pi G_{D}} \frac{(1+\lambda)b^{n+1}}{R^2(1+\lambda\frac{b^2}{R^2})}. \label{eq16}
\end{equation}
We impose the following conditions in our model:
\begin{itemize}
\item At the boundary of the star the interior solution should be matched with the Schwarzschild exterior solution, {\it i.e.},
\begin{equation}
e^{2\nu(r=b)} = e^{-2\mu(r=b)} = \left(1 - \frac{C}{b^{n-1}}\right), \label{eq17}
\end{equation}
where $C$ is a constant related to the mass of the star which is given by  $M=\frac{nA_{n}C}{16\pi G_{D}}$. Here $A_{n}=\frac{2 \pi^{(n+1)/2}}{\Gamma(n+1)/2}$. In four dimension ($D=4$), $C = 2~M$ and in five dimension ($D=5$), $C = 0.84848~MG_5$ where $G_5=GV_1$ and $V_1$ is the volume of extra space in five dimensions.

\item The radial pressure $p_{r}$ should vanish at the boundary of the star which gives,
\begin{equation}
\frac{\Psi_{z}(z_{b})}{\Psi(z_{b})} = - \frac{(n-1)(1+\lambda)}{2z_{b}} \label{eq18}
\end{equation}
where $z_{b}^2 =  (\lambda/(\lambda +1))(1 - b^2/R^2)$. From eq. (\ref{eq11}) one obtains
\begin{equation}
\frac{\psi{_z}}{\psi} = \frac{(\beta^2-1)}{\sqrt{(1-z^2)}} W
\label{eq19}
\end{equation}
where 
\[
W=\frac{\sin[(\beta-1)\zeta+\delta]-\sin[(\beta+1)\zeta
+\delta]}{(\beta+1)\cos[(\beta-1)\zeta+\delta]-(\beta-1)\cos[(\beta+1)\zeta +\delta]}. 
\]
Using  eqs. (\ref{eq18}) and (\ref{eq19}) we get
\begin{equation}
\tan\delta = \frac{\tau \cot\zeta{_b} - \tan(\beta\zeta{_b})}{1+\tau\cot\zeta{_b} \tan(\beta\zeta{_b})} \label{eq20}
\end{equation}
where $\tau =\frac{(n-1)(\lambda+1)-2\lambda \alpha}{\beta(1 + \lambda)(n-1)}$ and $\zeta{_b} = \cos ^{-1}z{_b}$.
\item As the radial pressure inside the star is positive, the condition $p_{r} \geq 0$ leads to the inequality
\begin{equation}
\frac{\Psi_{z}}{\Psi} \leq - \frac{(1+\lambda)(n-1)}{2z}. \label{eq21}
\end{equation}
\item Using eqs. (\ref{eq12})-(\ref{eq14}), the radial squared speed of sound is obtained which is given by
\begin{equation}
\frac{dp_{r}}{d\rho} = \frac{z(1-z^2)^2(\Psi_{z}/\Psi)^2 -(1-z^2)\Psi_{z}/\Psi) - \alpha \lambda z (1 - z^2)}{z(1-z^2)(\lambda+1)(n-1) + 4z} . \label{eq22}
\end{equation}
The variation of the tangential pressure with density is given by
\begin{equation}
\frac{d p_{t}}{d\rho} = \frac{d p_{r}}{d\rho} + {\alpha \lambda \over (1 + \lambda)} \bigg[ {(\lambda + 1)(1 - z^2) - 2 \over (\lambda + 1)(1 - z^2) + 4} \bigg]. \label{eq23}
\end{equation}
Now,  the parameters are so chosen that the causality conditions are not violated, {\it i.e.}, $\frac{dp_r}{d\rho},~ \frac{dp_{t}}{d\rho} \leq 1$ in the model.
\end{itemize}
The above constraints are used to obtain physically viable stellar models in the next section.

\section{Physical Analysis of Compact Objects}\label{Phy}

\begin{figure}[h!]
\begin{center}
\includegraphics{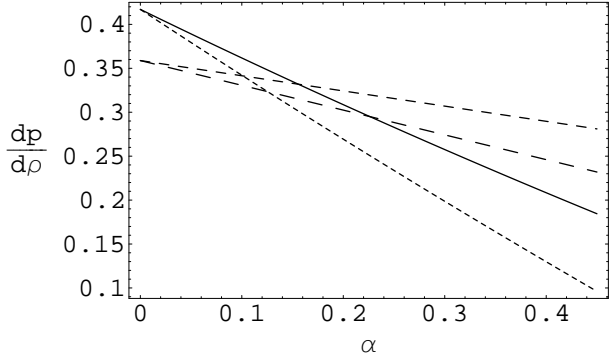}
\caption{\it {Variations of ($\frac{dp}{d\rho}$) at the centre of an anisotropic star with $\alpha$ for $\lambda = 53.34$, with $M = 1.435~M_{\odot}$, $b = 7.07$~km (SAX J 1808.4-3658). The solid and dotted lines represent the variation of $(\frac{dp_{r}}{d\rho})_{r=0}$ and $(\frac{dp_{t}}{d\rho})_{r=0}$ with $\alpha$ in four dimensions respectively. The dashed and long dashed lines represent  the variation of $(\frac{dp_{r}}{d\rho})_{r=0}$ and $(\frac{dp_{t}}{d\rho})_{r=0}$ in five dimensions respectively.}}
\label{fig1}
\end{center}
\end{figure}

\begin{figure}
\begin{center}
\includegraphics{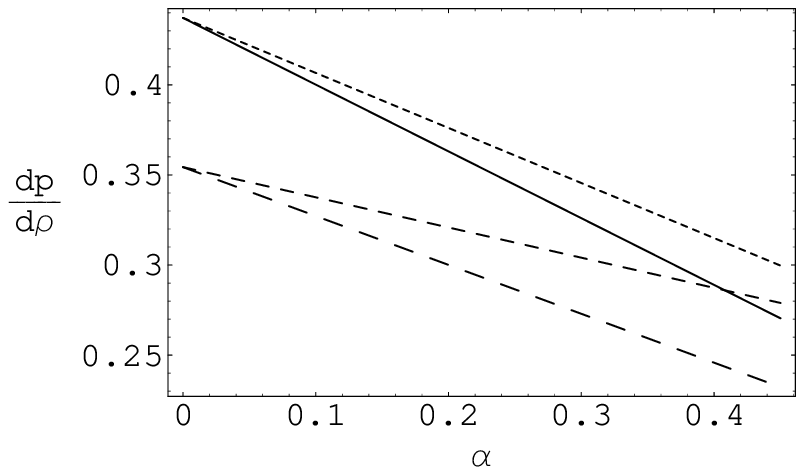}
\caption{\it {Variations of ($\frac{dp}{d\rho}$) at the centre of an anisotropic star with $\alpha$ for $\lambda = 53.34$, $M = 1.435~M_{\odot}$, $b = 7.07$~km (SAX J 1808.4-3658). The solid and dotted lines represent the variation of $(\frac{dp_{r}}{d\rho})_{r=b}$ and $(\frac{dp_{t}}{d\rho})_{r=b}$ with $\alpha$ in four dimensions respectively. The dashed and long dashed lines represent respectively the variation of $(\frac{dp_{r}}{d\rho})_{r=b}$ and $(\frac{dp_{t}}{d\rho})_{r=b}$ in five dimensions respectively.}}
\label{fig2}
\end{center}
\end{figure}

In this section we consider a higher dimensional space-time to determine the maximum mass of compact objects. We explore the effect of increasing the number of space-time dimensions in addition to anisotropy. The methodology adopted here is as follows :
 For a given  mass ($M$), radius ($b$), spheroidicity parameter ($\lambda$) and space-time dimensions ($D$), the factor $y = b^2/R^2$ can be determined from  eq. (\ref{eq16}). For a given  central or the surface density, the value of the geometrical parameter $R$  can be determined using  eq. (\ref{eq12}). Thereafter, the radius of  star $b = R\sqrt y$ and  mass $M$ can be determined using  eq. (\ref{eq16}).
  It may be mentioned here that for  a specific value of anisotropy parameter $\alpha$, the parameter $\delta$ is fixed.
 However, using  eqs. (\ref{eq22}) and (\ref{eq23}) one can show that for compact objects  with  same masses and radii might have different anisotropy for different equation of state (EoS). In sec. (\ref{Num}), it is shown that  EoS changes as one varies the anisotropy and space-time dimensions. It is evident from the  plot of variations of $\frac{d p_{r}}{d\rho}$ and $\frac{dp_{t}}{d\rho}$ with $\alpha$ in figs. \ref{fig1} and \ref{fig2} respectively.

It is also evident that the slope for  $\frac{dp}{d\rho}$ with anisotropy is decreases as the dimension is increased and moreover the value of  $\frac{dp}{d\rho}$  for five dimensions is less than that of four dimensions.  Thus the  EOS  of matter inside the star changes as the number of space-time dimensions are changed for the same set of values of the model parameters. For a given values of $\alpha$ and $\lambda$ one can determine $\delta$ from
 Eq. (\ref{eq20}). In the case of isotropic star ($\alpha = 0$), for a given $\lambda$ and $u_{iso}$ (isotropic compactness factor), we first calculate $y$ using eq. (\ref{eq16}) thereafter $\delta$ is determined from  eq. (\ref{eq20}). In the case of anisotropic star we employ same  $\delta$ to  determine 
  $y_{ani}$ for different $\alpha$.  Using  eq. (\ref{eq16}) for anisotropic compactness given by
   $$u_{ani} = \frac{M(b)}{b} = \frac{nA_{n}}{16\pi}\frac{(1 + \lambda)y_{ani}}{(1 + \lambda y_{ani})}$$ we probe the effect of anisotropy on the compactness of a star for different space-time dimensions.
We note that for vanishing anisotropy with $D=4$, the results are obtained by Karmakar {\em et al.} \cite{SK,SK1}.

\subsection{{\bf Numerical results}}\label{Num}

In this section  we consider two different compact objects of known masses as examples for the above purpose.

\begin{figure}[ht!]
\begin{center}
\includegraphics{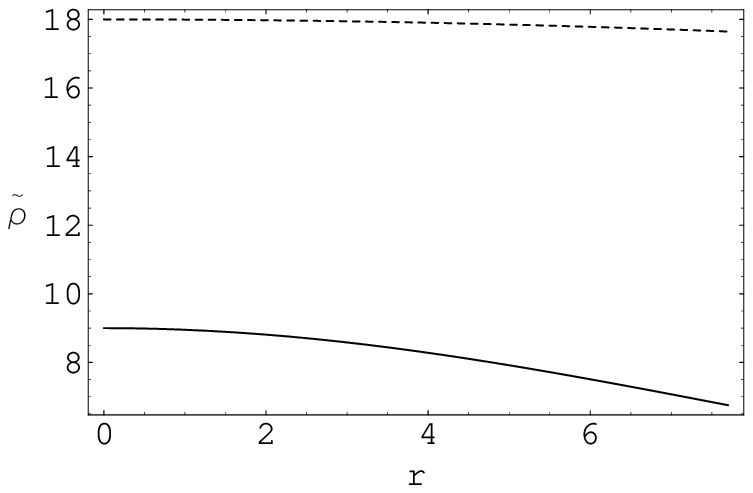}
\caption{\it {Radial variation of energy density $(\widetilde{\rho}=\rho R^2)$ interior to the star HER X-1 (Mass M=0.88~$M_{\odot}$, Radius=7.7~km). Solid line for $ D=4$ and dotted line for $D=5$ with anisotropy parameter $\alpha$=0.4 and $\lambda=2$.}}
\label{fig3}
\end{center}
\end{figure}

\begin{figure}
\begin{center}
\includegraphics{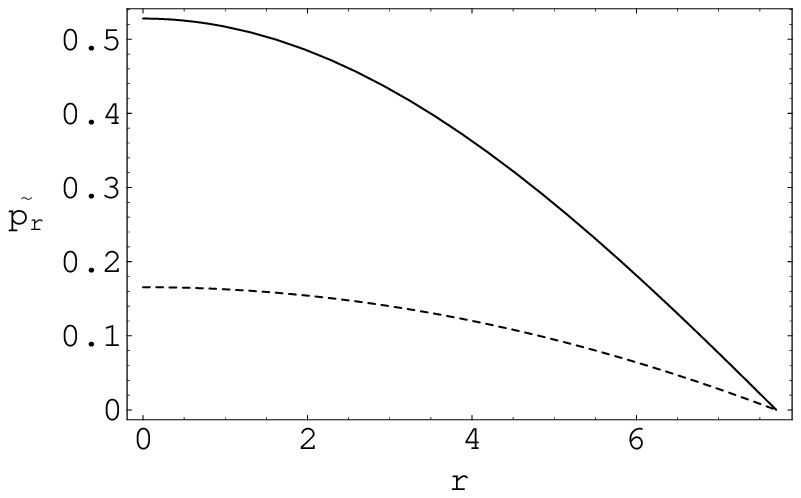}
\caption{\it {Variation of radial pressure ($\widetilde{p}_r = p_r R^2$) inside HER X-1 (Mass M=0.88~$M_{\odot}$, Radius=7.7~km). Solid line for $D=4$ and dotted line for $D=5$ for anisotropy parameter $\alpha$=0.4 and $\lambda=2$.}}
\label{fig4}
\end{center}
\end{figure}

\begin{figure}[ht!]
\begin{center}
\includegraphics{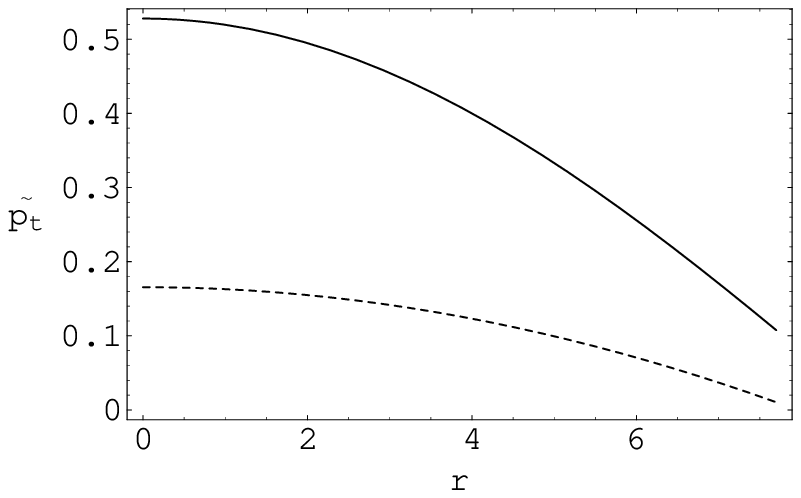}
\caption{\it {Radial variation of tangential pressure ($\widetilde{p}_{t} = p_{t} R^2$) inside  HER X-1 (Mass M=0.88~$M_{\odot}$, Radius=7.7~km). Solid line for $D=4$ and dotted line for $D=5 $ with  anisotropy parameter $\alpha$=0.4 and $\lambda=2$.}}
\label{fig5}
\end{center}
\end{figure}

{\bf Case I:} For the pulsar Her X-1 \citep{RS02} which has  mass $M = 0.88~ M_{\odot}$ where $M_{\odot}$ is the solar mass, radius $b = 7.7$~km,  in the framework of the space-time geometry considered here the  compactness factor is $u_{iso} = 0.1686$ when $\lambda = 2$ in four dimensions. Now as mentioned we determine $R$ using eq. (\ref{eq17})  for $\lambda=2$ in four and five dimensions which are $R = 20.2238$~km and $R=97.2474$~km respectively. It is now possible to study the radial variation  of energy density ($\rho$), radial pressure $(p_r)$ and transverse pressure $(p_{t})$  using eqs. (\ref{eq12}), (\ref{eq13}) and (\ref{eq14}) which are plotted in figs. (\ref{fig3}) - (\ref{fig5}) respectively for four (solid line) and five dimensions (dotted line).

In fig. (\ref{fig3}), we plot variation of  energy density $(\widetilde{\rho})$ inside the compact objects for a given $\alpha$ and $\lambda$ with different space-time dimensions. The radial pressure is found to increase with an increase in space-time dimensions (D). In the case of pressures plotted in  figs. (\ref{fig4}) and (\ref{fig5}), it is evident that both the radial and the tangential pressures decrease with increase of space-time dimensions. The rate of decrease of pressure is more when the dimension is less. The tangential pressure at the surface of the star is more in the case of lower dimension.

In Table-\ref{Tab1} we tabulated the calculated values of the compactness factor and mass of compact objects considering HER X-1 an anisotropic star. The observed mass is considered corresponding to isotropic compact object in  4 dimensions. In the case of  5-dimensions  mass of compact object is found  less than that of 4 dimensional mass. 
From Table-\ref{Tab1} it is evident from columns 3 and 4 that in four dimensions both  compactness factor ($u$) and mass ($M$) decreases with the increase of anisotropy $(\alpha)$. But in five dimensions both compactness factor ($u$) and mass ($M$) are found to  increase at first  with the increase of anisotropy parameter ($\alpha$), it attains a maximum value for a certain $\alpha$ then decreases as evident from  columns 7 and 8 respectively.

\begin{table}[ht!]
\begin{center}
\begin{tabular}{|l|l|l|l|l|l|l|r|}  \hline
\multicolumn{4}{|c|}{$D=4$, \; $\lambda = 2$, \; $b = 7.7~km$} &
\multicolumn{4}{|c|}{$D=5$, \; $\lambda = 2$, \; $b = 7.7~km$} \\ \hline
$\alpha$ & $\frac{y_{iso}}{1000}$ & $\frac{u}{100}$  & $\frac{M_{iso}}{10} $ & $\alpha$   & $\frac{y_{iso}}{1000}$ & $\frac{u}{100}$  & $\frac{M_{iso}}{10} $ \\ \hline
  0    &  145.0   & 16.86     & 8.8 $M_{\odot}$                &  0      &  6.27  & 2.18     & 1.14 $M_{\odot}$               \\ \hline
  
   \multicolumn{4}{|c|}{$D=4$, \; $\lambda = 2$, \; $b = 7.7~km$} &
\multicolumn{4}{|c|}{$D=5$,$\lambda = 2$, $b = 7.7~km$} \\ \hline                                                                          
  $\alpha$ & $\frac{y_{ani}}{1000}$ & $\frac{u_{ani}}{100}$  & $\frac{M_{ani}}{10}$ & $\alpha$ & $y_{ani}$ & $\frac{u_{ani}}{1000}$  & $\frac{M_{ani}}{10}$ \\ \hline
  0.2    &  128.1   & 15.30     & 8.0 $M_{\odot}$                &  0.2     &  7.14  &     2.48     & 1.29 $M_{\odot}$               \\ \hline
  0.4    &  95.3   & 12.01     & 6.3 $M_{\odot}$                 &  0.4     &  6.67   & 2.32     & 1.21 $M_{\odot}$               \\ \hline
  0.5    &  69.0   & 9.09     & 4.7 $M_{\odot}$                  &  0.5     &  5.80 & 2.02     & 1.05  $M_{\odot}$               \\ \hline
  0.6    &  32.2   & 4.54     & 2.4 $M_{\odot}$                 &  0.6     &  4.43  & 1.55     & 0.81 $M_{\odot}$               \\ \hline
\end{tabular}
\caption{\it {Compactness factor and mass calculated for different anisotropy parameter ($\alpha$) for space-time dimensions $D=4$ $\&$ $D=5$ }}
\label{Tab1}
\end{center}
\end{table}

{\bf Case II:} For a millisecond pulsar namely, SAX J 1808.4-3658  \cite{RS03} with  mass $M = 1.435~M_{\odot}$ and radius $b = 7.07$~km , it is found  that isotropic compactness $u_{iso} = 0.2994$ corresponds to   $\lambda = 53.34$. In this case also the compact star with the given mass and radius can be modelled  as an  anisotropic star. 

\begin{figure}[ht!]
\begin{center}
\includegraphics{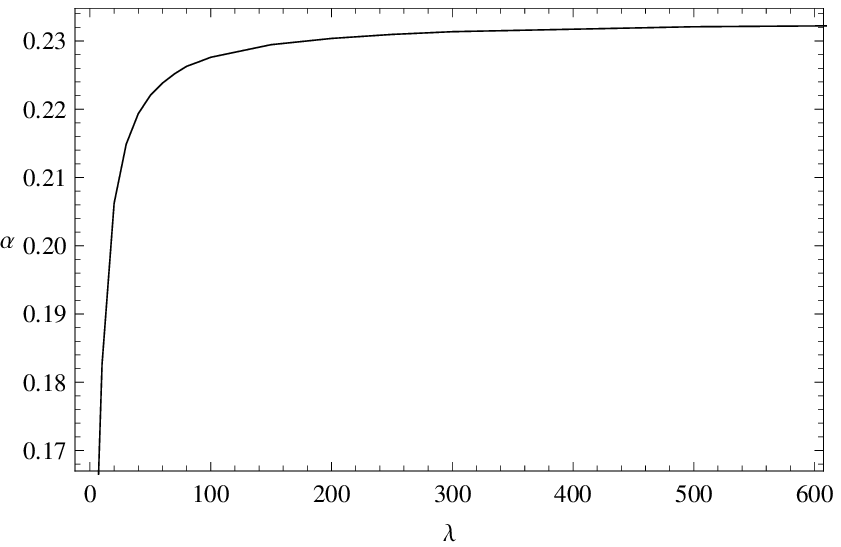}
\caption{\it {Variation of anisotropy parameter ($\alpha$) with spheroidicity parameter ($\lambda$) for a given mass and radius configuration.}}
\label{fig6}
\end{center}
\end{figure}

\begin{figure}
\begin{center}
\includegraphics{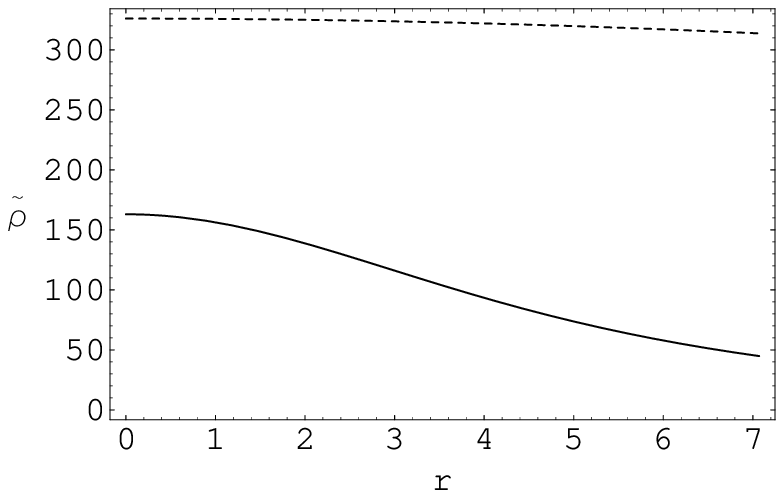}
\caption{\it {Radial variation of energy density $(\widetilde{\rho}=\rho R^2)$ interior to the star SAX J 1808.4-3658 (Mass M=1.435~$M_{\odot}$, Radius=7.07~km). Solid line for $D=4$ and dotted line for $ D=5 $ for anisotropy parameter $\alpha$=0.4 and $\lambda=53.34$.}}
\label{fig7}
\end{center}
\end{figure}

\begin{figure}
\begin{center}
\includegraphics{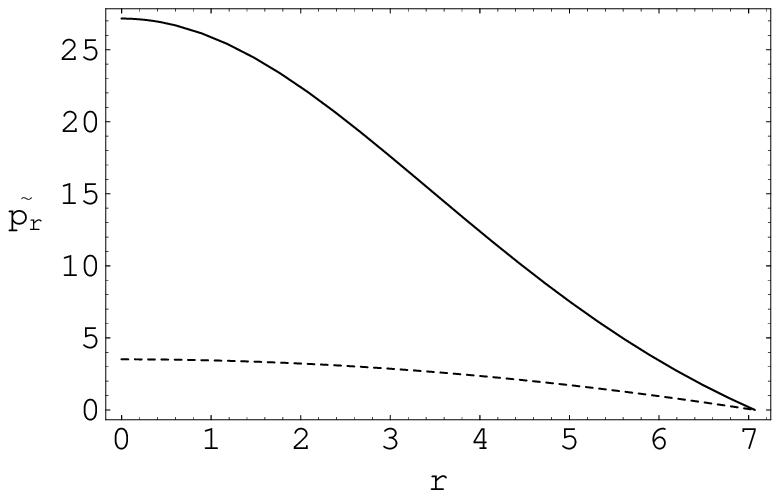}
\caption{\it {Variation of radial pressure $(\widetilde{p}_r = p_r R^2)$ interior to the star SAX J 1808.4-3658 (Mass M=1.435~$M_{\odot}$, Radius=7.07~km). Solid line for $D=4$ and dotted line for $D=5$ for anisotropy parameter $\alpha$=0.4 and $\lambda=53.34$.}}
\label{fig8}
\end{center}
\end{figure}

\begin{figure}
\begin{center}
\includegraphics{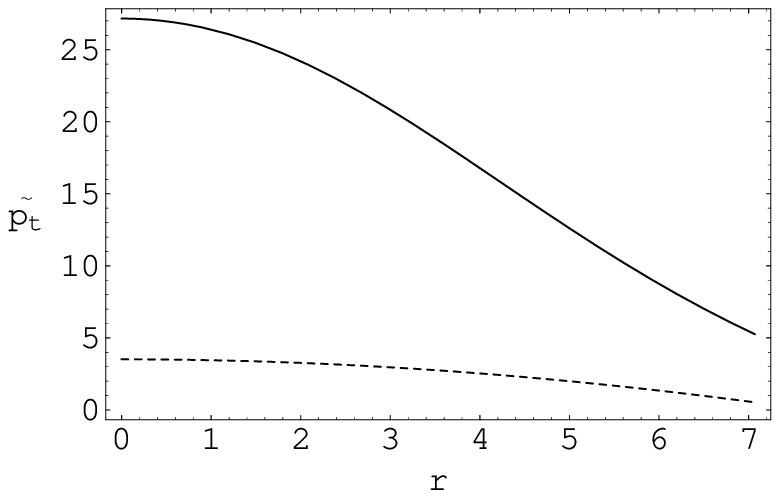}
\caption{\it {Radial variation of tangential pressure $(\widetilde{p}_{t} = p_{t} R^2)$ interior to the star SAX J 1808.4-3658 (Mass M=1.435~$M_{\odot}$, Radius=7.07~km). Solid line for $D=4$ and dotted line for $D=5$ for anisotropy parameter $\alpha$=0.4 and $\lambda=53.34$.}}
\label{fig9}
\end{center}
\end{figure}

It admits pressure anisotropy in the configuration for a  suitable combination of spheroidicity parameter ($\lambda$) and anisotropy parameter $(\alpha)$ in four dimensions as evident from fig. (\ref{fig6}). It is observed that the anisotropy parameter $\alpha$ increases with the increase of spheroidicity parameter $(\lambda)$ for a fixed mass of the star which we call isotropic mass ($M = 1.435~M_{\odot}$). It is also evident  that $\alpha$ increases first with the increase in $\lambda$ and at a large limiting value of $\lambda$ the anisotropy parameter  $\alpha$ attains a constant value. In this case we obtain $\alpha = 0.23267$ for $\lambda=600$. 
 The value of $R$ can be determined using eq. (\ref{eq17}) for $\lambda=53.34$ which gives  $R =43.245$~km and $R=270.059$~km in four and five dimensions respectively. We plot the radial variation of  energy density ($\rho$), radial pressure $(p_r)$ and transverse pressure $(p_{t})$ in four and five dimensions in figs. (\ref{fig7})-(\ref{fig9}) respectively. 
 In fig. (\ref{fig7}), we plot the variation of energy density $(\widetilde{\rho})$ with radial distance and found that energy density is more in five dimensions than that in four dimensions. A star of same radius accommodates more mass in the case of higher dimensions. However radial and transverse pressures are found 
 to have  lower values in higher dimensions than that in four dimensions which is  evident from figs. (\ref{fig8}) and (\ref{fig9}) respectively.
 In Table-\ref{Tab2},  the values of $y$, $u$ and $M$ are given considering  an isotropic star ($\alpha=0$) and also for an   anisotropic  ($\alpha \neq 0$) stellar configuration both in 4 and 5-dimensions.
 It is   evident from columns 3 and 4 that in four dimensions both compactness factor ($u$) and the corresponding mass of a star ($M$) first increases with an increase in anisotropy $(\alpha)$, which attains  a   maximum value and thereafter decreases. But in five dimensions both compactness factor and mass of a compact object are found to  decrease with the increase of anisotropy parameter ($\alpha$) as evident from  columns 7 and 8 of Table-\ref{Tab2}. It is also evident that for the same anisotropy, compactness of a star is found to decrease significantly if the space-time dimensions are increased.  We note that there is a limiting value of $\alpha$ above which $y=b^2/R^2$ becomes zero or negative which is non-physical. 
 
 It is noted that for HER X-1, a physically realistic stellar model is obtained with a maximum value of $\alpha$ which are $0.665$ and $0.79$ for $D=4$ and $D=5$ respectively. In the case of SAX J 1808.4-3658, however, a physically realistic stellar model is permissible with a maximum $\alpha$ which are $0.65$ and $0.49$ for $D=4$ and $D=5$ respectively.

\begin{figure}
\begin{center}
\includegraphics{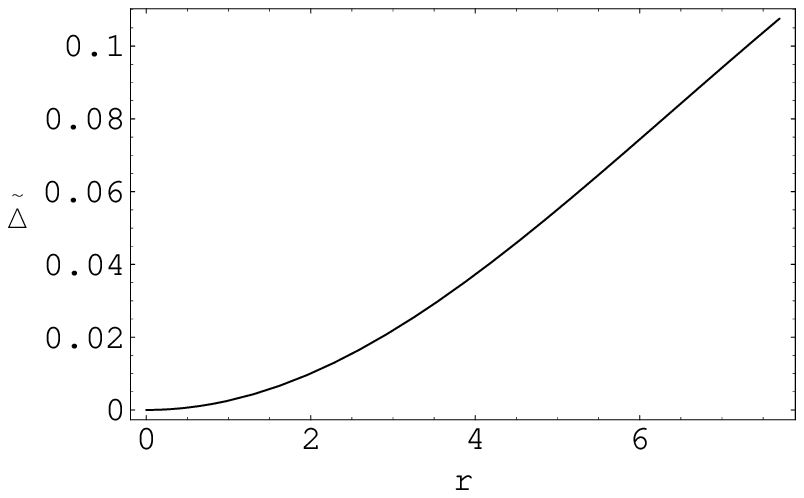}
\caption{\it {Radial variation of $\widetilde{\Delta}=\Delta~R^2$ interior to the star HER X-1 with $u_{iso}=0.1686$, $\lambda=2$, $\alpha=0.4$ and $D=4$.}}
\label{fig10}
\end{center}
\end{figure}

\begin{figure}
\begin{center}
\includegraphics{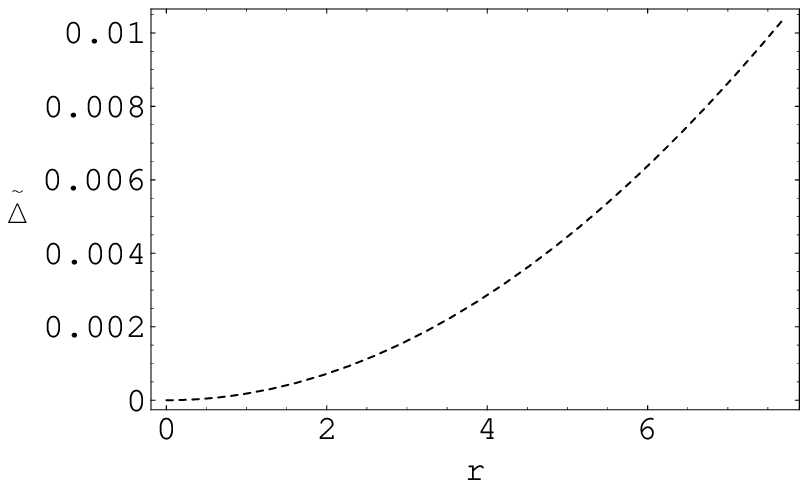}
\caption{\it {Radial variation of $\widetilde{\Delta}=\Delta~R^2$ interior to the star HER X-1 with $u_{iso}=0.1686$, $\lambda=2$, $\alpha=0.4$ and $D=5$.}}
\label{fig11}
\end{center}
\end{figure}

\begin{figure}
\begin{center}
\includegraphics{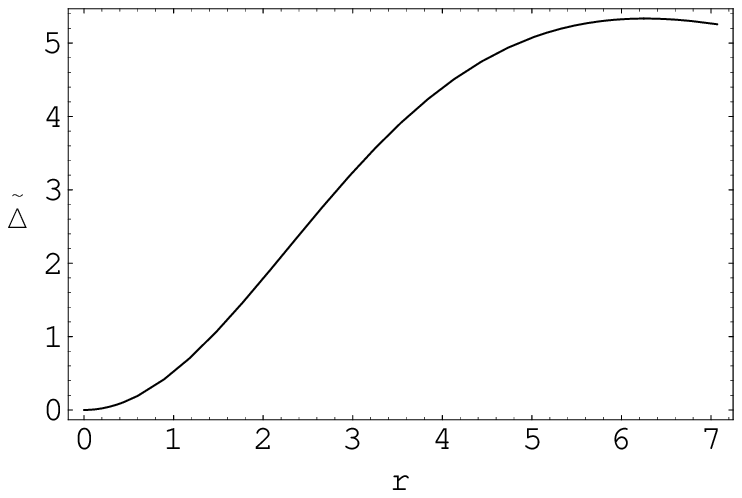}
\caption{\it {Radial variation of $\widetilde{\Delta}=\Delta~R^2$ interior to the star SAX J 1808.4-3658 with $u_{iso}=0.2994$, $\lambda=53.34$, $\alpha=0.4$ and $D=4$.}}
\label{fig12}
\end{center}
\end{figure}

\begin{figure}
\begin{center}
\includegraphics{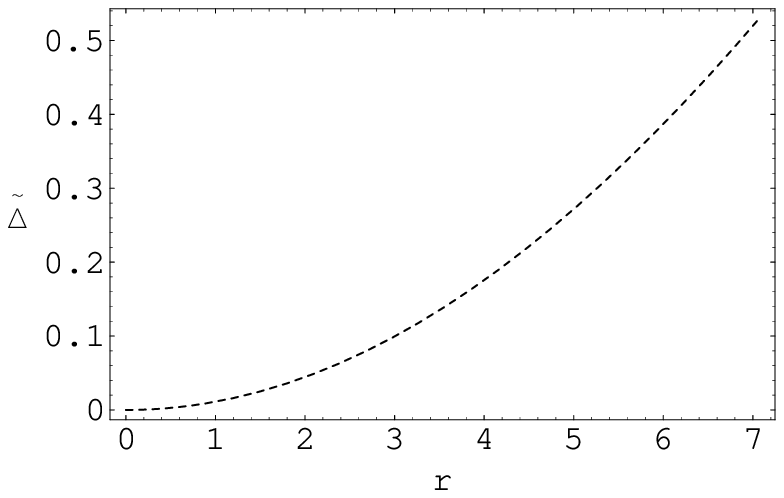}
\caption{\it {Radial variation of $\widetilde{\Delta}=\Delta~R^2$ interior to the star SAX J 1808.4-3658 with $u_{iso}=0.2994$, $\lambda=53.34$, $\alpha=0.4$ and $D=5$.}}
\label{fig13}
\end{center}
\end{figure}

\begin{table}
\begin{center}
\begin{tabular}{|l|l|l|l|l|l|l|r|}  \hline
\multicolumn{4}{|c|}{$D=4$, $b = 7.07~km$} &
\multicolumn{4}{|c|}{$D=5$, $b = 7.07~km$} \\ \hline
 $\alpha$ &  $\frac{y_{iso}}{1000}$ & $\frac{u}{100}$ & $\frac{M_{iso}}{10}$ & $\alpha$ & $\frac{y_{iso}}{1000}$ & $\frac{u}{10}$ & $\frac{M_{iso}}{10}$ \\ \hline
  0     &   26.7   &  29.94   &  1.435$M_{\odot}$               &    0  & 0.68   & 4.23    & 0.203                \\ \hline
  
  \multicolumn{4}{|c|}{$D=4$,  \; $b = 7.07~km$} &
\multicolumn{4}{|c|}{$D=5$,  \; $b = 7.07~km$} \\ \hline

  $\alpha$ & $\frac{y_{ani}}{1000}$ & $\frac{u_{ani}}{100}$  & $M_{ani}$ & $\alpha $ & $\frac{y_{ani}}{1000}$ & $\frac{u_{ani}}{100}$  & $M_{ani}$ \\ \hline
  0.2    &   26.8   & 29.97   &  1.437 $M_{\odot}$               &    0.2   & 0.62   & 3.84    & 0.184 $M_{\odot}$                \\ \hline
  0.4    &   23.9   & 28.58   &  1.368 $M_{\odot}$              &    0.3   & 0.49   & 3.06    & 0.147 $M_{\odot}$                \\ \hline
  0.5    &   19.5   &  26.00   &  1.246 $M_{\odot}$               &    0.4   & 0.28   & 1.77    & 0.085 $M_{\odot}$                \\ \hline
  0.6    &   9.0   &  17.60   &  0.844  $M_{\odot}$              &    0.45  & 0.14   & 0.89    & 0.043 $M_{\odot}$                \\ \hline
\end{tabular}
\caption{\it {Compactness  factor and mass for different anisotropy ($\alpha$) in $ D=4 $ $\&$  $D=5$ dimensions with $\lambda=53.34$}}
\label{Tab2}
\end{center}
\end{table}


\begin{table}
\begin{center}
\begin{tabular}{|c|c|c|c|c|c|c|}  \hline\hline
$D$ & $\alpha$ & $\delta$ &  $y_{max}$ & $(\frac{M}{b})_{max}$ & $(Z_s)_{max}$ & $M_{max}$   \\ \hline\hline
4   & 0        & 1.71374  & 0.42341    & 0.34390              & 0.78970      & 2.3315 $M_{\odot}$  \\
    & 0.5      & 1.65996  & 0.47337    & 0.36474              & 0.92267      & 2.4728 $M_{\odot}$   \\  \hline
5   & 0        & 1.66216  & 0.14400    & 0.39514              & 1.18368      & 2.6790 $M_{\odot}$   \\
    & 0.5      & 1.65254  & 0.16517    & 0.43881              & 1.85850      & 2.9750 $M_{\odot}$   \\ \hline
6   & 0        & 1.51764  & 0.03560    & 0.20882              & 0.31039      & 1.4157 $M_{\odot}$   \\
    & 0.5      & 1.53075  & 0.04521    & 0.26051              & 0.44491      & 1.7662 $M_{\odot}$   \\ \hline
\end{tabular}
\caption{Maximum Mass configurations for a star of radius $10~km.$ with $\lambda=2$}
\label{Tab3}
\end{center}
\end{table}

\begin{table}
\begin{center}
\begin{tabular}{|c|c|c|c|c|c|c|}  \hline\hline
$D$ & $\alpha$ & $\delta$ &  $\frac{y_{max}}{10}$ & $(\frac{M}{b})_{max}$ & $Z_s|_{max}$ & $M_{max}$   \\ \hline\hline
4   & 0        & 1.718  & 3.728    & 0.352              & 0.838      & 2.386 $M_{\odot}$   \\
    & 0.5      & 1.654  & 4.283    & 0.375             & 0.999      & 2.541 $M_{\odot}$  \\  \hline
5   & 0        & 1.681  & 1.294    & 0.439              & 1.869     & 2.978 $M_{\odot}$   \\
    & 0.5      & 1.665  & 1.505    & 0.489              & 5.602      & 3.312 $M_{\odot}$   \\ \hline
6   & 0        & 1.546  & 0.410    & 0.306              & 0.608      & 2.075 $M_{\odot}$   \\
    & 0.5      & 1.557  & 0.502    & 0.365              & 0.928      & 2.477 $M_{\odot}$   \\ \hline

\end{tabular}
\caption{\it {Maximum Mass configurations for a star of radius $10~km$ with $\lambda=3$.}}
\label{Tab4}
\end{center}
\end{table}

\begin{table}
\begin{center}
\begin{tabular}{|c|c|c|c|c|c|c|}  \hline\hline
$D$  & $n$  &          & $\frac{y_{max}}{1000}$ & $(\frac{M}{b})_{max}$ & Maximum Mass         \\
                &      & $\alpha$ &           &                       & $M_{max}(M_{\odot})$ \\ \hline\hline
 4              &  2   &     0               &   25.19 & 0.3615                &   2.4510             \\
                &      &     0.5             &   45.02 & 0.3905                &   2.6475             \\  \hline
 5              &  3   &     0               &   7.82 & 0.5222                &   3.5407             \\
                &      &     0.5             &   12.02 & 0.5874                &   3.9824             \\ \hline
 6              &  4   &     0               &   3.09 & 0.4994                &   3.3855             \\
                &      &     0.5             &   4.58 & 0.5816                &   3.9428             \\ \hline
 7              &  5   &     0               &   1.09 & 0.3080                &   2.0881             \\
                &      &     0.5             &   1.80 & 0.3903                &   2.6461             \\ \hline
 8              &  6   &     0               &   0.06 & 0.0238                &   0.1614             \\
                &      &     0.5             &   0.45 & 0.9716                &   0.6587             \\ \hline
\end{tabular}
\caption{\it {Maximum Mass configurations for a star of radius $10~km$ with $\lambda=100$.}}
\label{Tab5}
\end{center}
\end{table}

\section{Maximum mass and Surface Red-shift}\label{Max}

In this section  maximum mass of a class of isotropic and anisotropic stars in four ($D=4$) and in higher $(D >  4)$ dimensions will be explored. In determining the maximum mass of a compact object in higher dimensions we follow a technique  adopted in Ref. (\cite{SK,SK1}). 
\begin{itemize}
\item The squared speed of sound  should satisfy an inequality ($\frac{d p{_r}}{d\rho} \leq 1$) inside the compact object for causality. It decreases away from the centre thus we consider squared speed of sound maximum at the centre which leads to 
\begin{equation}
\frac{\psi_{z}}{\psi}|_{zo} \geq \frac{(1+\lambda)}{2\sqrt{\lambda}}
\left[\sqrt{\lambda+1} - \sqrt{(4 n +13)\lambda+1 + \frac{4 \alpha \lambda^2}{\lambda + 1}}\right]. \label{eq24}
\end{equation}
Using (\ref{eq18}) and (\ref{eq24}), one can determine the limiting value of $\delta$ which is a function of  $\alpha$ for given  values of $\lambda$ and $D$.
\item Corresponding to the limiting value of $\delta$,   a  maximum value for $ y = b^2/R^2$ can be determined using eq. (\ref{eq19}).
\item  From  eq. (\ref{eq15}) the compactness of a compact star in higher dimension can be determined which is given by
\begin{equation}
u = \frac{M(b)}{b} =\frac{nA_{n}}{16\pi} \frac{(1+\lambda)}{(\lambda + \frac{1}{y})}.\label{eq25}
\end{equation}
Thus the maximum value of $y$ corresponds to the maximum compactness of a stellar configuration. The maximum surface red-shift ($(Z_s)_{max}$) is given by :
\begin{equation}
(Z_s)_{max}=\left(1-2u_{ani}\right)^{-1/2} -1.\label{eq26}
\end{equation}
\end{itemize}

\begin{figure}
\begin{center}
\includegraphics{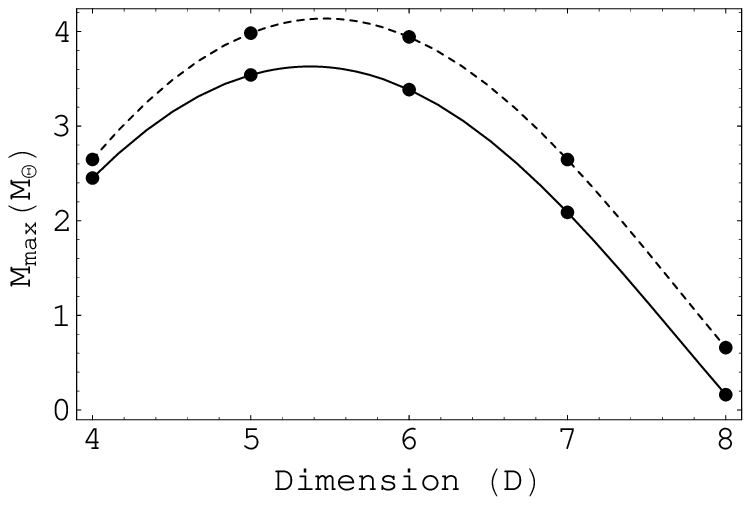}
\caption{\it {Variation of Maximum mass $(M_{max})$ with Dimensions ($D$) for a compact object with radius $b = 10$~km and $\lambda=100$. Solid curve for $\alpha=0$ and dotted curve for $\alpha=0.5$.}}
\label{fig14}
\end{center}
\end{figure}

Thus, in the above it is noted that once the maximum compactness of a star is known the corresponding maximum mass of the anisotropic star can be determined for a given radius or surface density. In Tables-\ref{Tab3}, \ref{Tab4} and \ref{Tab5} we have tabulated the maximum mass of stars,  surface red-shift for $\lambda=2$, $3$ and $100$ respectively in four and higher dimensions. From Tables-\ref{Tab3} and \ref{Tab4}, it is evident that  both the surface red-shift and maximum mass are found to increase with the increase of anisotropy parameter $(\alpha)$ for a given dimension. However surface red-shift and maximum mass both  increases  when the  space-time dimensions are increased with or without anisotropy. 
We  note that for  spheroidicity parameter $\lambda=100$, the maximum compactness of an isotropic star is $0.3615$ which is same as that obtained in Ref. \citep{SK}. We note that the maximum mass of a compact star first increases with the increase of dimensions, attains a maximum value in between $D=5$ and $D=6$ (if fractal dimensions  exist), thereafter it decreases which is evident in  Table-\ref{Tab5}. In  fig-(\ref{fig14}) we plot the variation of maximum mass with dimension. 
From Table-\ref{Tab5}, it is evident that maximum  compactness factor of a compact object may exceed $0.5$ in  higher dimensions. In Kaluza-Klein gravity similar limiting value of  compactness of a higher dimensional star admitting  compactness more than  $0.5$ without a black hole was reported in the literature \citep{JP}. However if one restricts the compactness to a value less then $0.5$, then the maximum allowed value of $\lambda$ found in this case is $10$ in isotropic star. In Tables-\ref{Tab3}, \ref{Tab4} and \ref{Tab5}, it is evident that
 for a given value of $\lambda$ there exists a upper limit of space-time  dimensions for a physically viable model. It is evident that a compact object with spheroidicity parameters $\lambda=2$ and $3$ can be accommodated consistently in $D\leq6$ and for a large value say, $\lambda = 100$ it can be accommodated in $D\leq8$. 

\subsection{Equation of State (EoS)}

Using the above model parameters we  plot the radial variation of density and radial pressure vide  eqs. (\ref{eq12}) and (\ref{eq13}). However, it may be pointed out here that  an analytic function of pressure with density in known form cannot be obtained here because of complexity of the equations. We study numerically to obtain a best fitted  relation between the energy density $(\rho)$ and radial pressure ($p$) which are presented in  Tables-\ref{Tab6} and \ref{Tab7}.  Theoretically a convenient way of expressing EoS is obtained from energy per unit mass of the fluid which is a function of energy density ($u$) and entropy ($S$) respectively. From the first law of thermodynamics,
\begin{equation}
du=- \, p \, d\left(\frac{1}{\rho}\right)+T \, dS.\label{eq27}
\end{equation}

For the description of the fluid flow  pressure and  temperatures are given by
\begin{equation}
p=\rho^2\frac{\partial u}{\partial \rho}|_{S}, \;\;\;\;T=\frac{\partial u}{\partial S}|_{\rho}.\label{eq28}
\end{equation}
In the case of production of entropy through dissipative processes we restrict to adiabatic flows only. In the  isentropic case $\frac{dS}{dt}=0$, entropy remains constant. Therefore, the energy density becomes a one parameter function. Consequently eq. (\ref{eq27}) is equivalent to
\begin{equation}
p=p(\rho).\label{eq29}
\end{equation}
It is evident that the EoS obtained  (in Table-\ref{Tab6} and \ref{Tab7}) numerically using the eqs. (\ref{eq12}) and (\ref{eq13}) is not  linear relation of the form $ p = \omega \rho$ where $\omega$ is a constant. We found that the models may be fitted with linear, quadratic even with higher order polynomial function in $\rho$. We determine here two probable EoS in isotropic and anisotropic case  both in $D=4$ and $D=5$. The EoS obtained here are found to have similar to that recently considered by Maharaj and Mafa Takisa \cite{Maharaj1}. From Tables-\ref{Tab6} and \ref{Tab7} we note that equation of state becomes softer in higher dimensions. Using suitable choice of $\lambda$ and $\alpha$ in some compact objects  it may be possible to fit the equation of state with  $p_r=\frac{1}{3}(\rho-4B)$, where $B$ is the Bag constant in MIT Bag model for strange matter. This aspect of the model will be taken up elsewhere.

\section{Discussion}\label{Dis}

In this paper we study  compact objects in hydrostatic equilibrium making use of an  alternative approach considered by Mukherjee {\it et. al.} \citep{Mukherjee}. The interior geometry is described by Vaidya-Tikekar metric  both in four and in higher dimensions. 
The radial variation of pressure ($p_r$) for HER X-1 and SAX J 1808.4-3658 are shown in figs. (\ref{fig4}) and (\ref{fig8}) respectively assuming an anisotropic distribution of fluid. Radial variation of transverse pressure ($p_t$) for the two stars mentioned here are also shown in figs. (\ref{fig5}) $\&$ (\ref{fig9}). It is evident from the figures that both $p_r$ and $p_{\perp}$ decreases from the centre to the surface of the stars both in four and five dimensions. This type of variation is also found for anisotropic stellar models as obtained by Chaisi and Maharaj \citep{Chaisi} and Sharma {\it et. al.} \citep{RS03} for four dimensional space-time geometry. The radial variation of anisotropy in pressure ($\Delta$) in case of HER X-1 are plotted in figs. (\ref{fig10}) and (\ref{fig11}) in $D=4$ and $D=5$ respectively. The radial variation of anisotropy in pressure ($\Delta$) in case of SAX J 1808.4-3658 are plotted in figs. (\ref{fig12}) and (\ref{fig13}) in $D=4$ and $D=5$ respectively. The variation in tangential pressure is physically acceptable. Since during the quasi-equilibrium contraction of a massive body, conservation of angular momentum leads to a high value for transverse pressure at the central region of the star. To incorporate the effect of dimensions on the maximum mass of a star, we obtain a class of relativistic solution in spheroidal space-time in Vaidya-Tikekar model with higher dimensions. The solution is then employed to estimate the maximum mass of a star in higher dimensions. The maximum mass of a isotropic star and that in the presence of anisotropy are also discussed in \citep{SK1} and \citep{SK} respectively. We recover the maximum mass obtained in four dimensions in isotropic case $(2.45~M_{\odot})$ and in presence of anisotropy $(2.8~M_{\odot})$ for $\lambda=100$. We also note that the maximum mass increases with the increase of space-time dimensions (D) which is maximum in between $D=5$ and $D=6$, thereafter it decreases. It is found that $M_{max}=3.54~M_{\odot}$ when $\alpha=0$ and $M_{max}=3.9824~M_{\odot}$ when $\alpha=0.5$ in 5-dimensions as shown in Table-(\ref{Tab5}). It is also noted that in higher dimensions the maximum mass of a anisotropic star is greater than an isotropic star also. From figs. (\ref{fig10})- (\ref{fig13}), it is evident that at the centre of the star anisotropy $(\Delta)$ vanishes both in D=4 and D=5, whereas at the surface it attains a maximum value. Though the nature of radial variation of $\Delta$ is same, $\Delta$ picks up lower values in higher dimensions. We also note that the surface red-shift has greater value in case of anisotropic star than isotropic one. Also surface red-shift increases with dimensions first, attains a maximum value and then decreases. It picks up a maximum value when mass of the star attains its maximum {\it i.e.} we note a correlation between maximum mass of a star and its surface red-shift.

\begin{table}[ht!]
\begin{center}
\begin{tabular}{|c|c|c|c|c|c|} \hline
\multicolumn{1}{|c|}{Star}&\multicolumn{1}{|c|}{Mass}&\multicolumn{1}{|c|}{Size}&\multicolumn{1}{|c|}{$\lambda$} & \multicolumn{1}{|c|}{$\alpha$}&\multicolumn{1}{|c|}{Equation of State} \\ \hline
HER X1 &                  &      & 2   & 0   & $p=0.1747\rho-0.1518$              \\ \cline{6-6}
       & 0.88$M_{\odot}$  &7.7   &     &     & $p=-0.0067\rho^2+0.189\rho-0.156$  \\ \cline{5-6}
       &                  &      &     & 0.3 & $p=0.1466\rho-0.1272$              \\ \cline{6-6}
       &	                &      &     &	   & $p=-0.0125\rho^2+0.174\rho-0.142$  \\ \hline
SAX J  &                  &      &     & 0   & $p=0.2309\rho-0.299$               \\ \cline{6-6}
       & 1.435$M_{\odot}$ &7.07  &53.34&     & $p=-0.0041\rho^2+0.258\rho-0.336$  \\ \cline{5-6}
SS1    &                  &      &     & 0.3 & $p=0.1527\rho-0.1892$              \\ \cline{6-6}
			 &                  &      &     &	   & $p=-0.0061\rho^2+0.193\rho-0.244$  \\ \hline
SAX J  &                  &      &     & 0   & $p=0.2627\rho-0.4393$              \\ \cline{6-6}
       & 1.323$M_{\odot}$ &6.55  &5    &     & $p=-0.0025\rho^2+0.2800\rho-0.465$ \\ \cline{5-6}
SS2    &                  &      &     & 0.3 & $p=0.201\rho-0.3289$               \\ \cline{6-6}
       &                  &      &     &     & $p=-0.0057\rho^2+0.239\rho-0.387$  \\ \hline
\end{tabular}
\caption{\it {Equation of state for different stellar models in 4-dimensions.}}
\label{Tab6}
\end{center}
\end{table}
\begin{table}
\begin{center}
\begin{tabular}{|c|c|c|c|c|c|}  \hline
\multicolumn{1}{|c|}{Star}&\multicolumn{1}{|c|}{Mass}&\multicolumn{1}{|c|}{Size}&\multicolumn{1}{|c|}{$\lambda$} & \multicolumn{1}{|c|}{$\alpha$}&\multicolumn{1}{|c|}{Equation of State} \\  \hline
HER X1 &                  &      & 2   & 0   & $p=0.2654\rho-0.0695$              \\ \cline{6-6}
       & 0.88$M_{\odot}$  &7.7   &     &     & $p=0.2002\rho^2+0.158\rho-0.055$   \\ \cline{5-6}
       &                  &      &     & 0.3 & $p=0.248\rho-0.065$                \\ \cline{6-6}
       &	                &      &     &	   & $p=0.165\rho^2+0.159\rho-0.053$    \\ \hline
SAX J  &                  &      &     & 0   & $p=0.1917\rho-0.1121$              \\ \cline{6-6}
       & 1.435$M_{\odot}$ &7.07  &53.34&     & $p=0.0364\rho^2+0.146\rho-0.098$   \\ \cline{5-6}
SS1    &                  &      &     & 0.3 & $p=0.1646\rho-0.0962$              \\ \cline{6-6}
			 &                  &      &     &	   & $p=0.0234\rho^2+0.1352\rho-0.087$  \\ \hline
SAX J  &                  &      &     & 0   & $p=0.2277\rho-0.1673$              \\ \cline{6-6}
       & 1.323$M_{\odot}$ &6.55  &5    &     & $p=0.045\rho^2+0.1572\rho-0.140$   \\ \cline{5-6}
SS2    &                  &      &     & 0.3 & $p=0.2046\rho-0.1503$              \\ \cline{6-6}
       &                  &      &     &     & $p=0.0333\rho^2+0.1524\rho-0.13$   \\ \hline
\end{tabular}
\caption{\it {Equation of state for different stellar models in 5-dimensions.}}
\label{Tab7}
\end{center}
\end{table}

{\it Acknowledgement}

The authors would like to thank IUCAA Resource Centre (IRC) at Physics Department, North Bengal University, Siliguri for providing faciities to carry out the researh work. BCP would like to thank University Grants Commission (UGC), New Delhi for awarding a Major Research Project (F.42-783/(2013)SR) and TWAS-UNESCO for Visiting Associateship.


\begin{thebibliography}{99}

\bibitem[\protect\citeauthoryear{Kaluza}{1921}]{k1} Kaluza, T.: {\it Sitz. Preuss. Acad. Wiss.} {\bf F1} 966 (1921)

\bibitem[\protect\citeauthoryear{Klein}{1926}]{k2}   Klein, O.:   {\it A. Phys.} {\bf 37} 895 (1926).

\bibitem[\protect\citeauthoryear{Chodos \& Detweiler}{1980}]{cd}  Chodos, A., Detweiler, S.: {\it Phys. Rev.} {\bf D 21} 2167 (1980)

\bibitem[\protect\citeauthoryear{Shafi \& Wetterich}{1987}]{cd1}  Shafi, Q., Wetterich, C.: {\it Nucl. Phys.} {\bf B 289} 787 (1987)

\bibitem[\protect\citeauthoryear{Wetterich}{1982}]{cd2} Wetterich, C.: {\it Phys. Lett.} {\bf B 113} 377 (1982)

\bibitem[\protect\citeauthoryear{Accetta et al.}{1986}]{cd3} Accetta, F.S., Gleicer, M., Holman, R., Kolb, E.W.:  {\it Nucl. Phys.} {\bf B 276} 501 (1986)

\bibitem[\protect\citeauthoryear{Lorentz-Petzold}{1988}]{lp} Lorentz-Petzold, D.: {\it Prog. Theor. Phys.} {\bf B 78} 969 (1988){\it Mod. Phys. Lett.} {\bf A 3} 827 (1988)


\bibitem[\protect\citeauthoryear{Paul \& Mukherjee}{1990}]{cd6} Paul, B.C., Mukherjee, S.: {\it Phys. Rev.} {\bf D 42} 2595 (1990)

\bibitem[\protect\citeauthoryear{Green \& Schwarz}{1984}]{str} Green, M.B., Schwarz, J.H.: {\it Phys. Lett.} {\bf B 149} 117 (1984)
\bibitem[\protect\citeauthoryear{Green \& Schwarz}{1985}]{str1} Green, M.B., Schwarz, J.H.: {\it Phys. Lett.} {\bf B 151} 21 (1985)
\bibitem[\protect\citeauthoryear{Candelas et al.}{1985}]{str2} Candelas, P., Horowitz, G., Strominger, A., Witten, E.: {\it Nucl. Phys. } {\bf B 258} 46 (1985)
\bibitem[\protect\citeauthoryear{Witten}{1995}]{str3} Witten, E.: {\it Nucl. Phys.} {\bf B 443} 85 (1995).
\bibitem[\protect\citeauthoryear{Randal \& Sundrum}{1999}]{rs} Randal, L., Sundrum, R.: {\it Phys. Rev. Lett.} {\bf 83} 3370 (1999);{\it Phys. Rev. Lett.} {\bf 83} 4690 (1999)


\bibitem[\protect\citeauthoryear{Chodos  \& Detweiler}{1982}]{RN}  Chodos, A., Detweiler, S.: {\it Gen. Rel. Grav.} {\bf 14} 879 (1982)
\bibitem[\protect\citeauthoryear{Gibbons \& Wiltshire}{1986}]{RN1} Gibbons G. W., Wiltshire, D. L. : {\it Ann. Phys. } (N.Y) {\bf 167} 201 (1986)
\bibitem[\protect\citeauthoryear{Myers \& Perry}{1986}]{Mperry}  Myers, R. C.,  Perry, M. J.: {\it Ann. Phys.} (N.Y)  {\bf 172} 304 (1986)

\bibitem[\protect\citeauthoryear{Mazur}{1987}]{Kerr} Mazur, P.O. : {\it Math. Phys.  } {\bf 28} 406 (1987)
\bibitem[\protect\citeauthoryear{Xu}{1988}]{Kerr1}  Xu, D.: {\it Class. Quantum Grav.} {\bf 5} 871 (1988)

\bibitem[\protect\citeauthoryear{Myers}{1986}]{Myers} Myers, R. C. : {\it Phys. Rev.} {\bf  D 35} 455 (1986)
\bibitem[\protect\citeauthoryear{Sokolowski \& Carr}{1986}]{Sokolowski}  Sokolowski, L.,  Carr, B.: {\it Phys. Lett.} {\bf B 176 } 334 (1986)
\bibitem[\protect\citeauthoryear{Iyer \& Vishveshwara}{1989}]{Iyer}  Iyer, B., Vishveshwara,C. V. : {\it Pramana J. Phys.} {\bf 32} 749 (1989)
\bibitem[\protect\citeauthoryear{Shen \& Tan}{1989}]{STan} Shen T., Tan, Z.: {\it Phys. Lett. } {\bf A 142} 341 (1989)
\bibitem[\protect\citeauthoryear{Paul}{2001}]{bcp} Paul, B.C.: {\it Class. Quantum Grav.} {\bf  18} 2311 (2001)
\bibitem[\protect\citeauthoryear{Liddle et al.}{1990}]{LiddleAR} Liddle, A. F.,  Moorhouse, R. G.,  Henriques, A. B. : {\it Class. Quantum Grav.} {\bf 7} 1009 (1990)
\bibitem[\protect\citeauthoryear{Herrera \& Santos}{1997}]{Herrera} Herrera, L.,  Santos,  N. O. : {\it Phys. Rep.} {\bf 286} 53 (1997)
\bibitem[\protect\citeauthoryear{Tikekar \& Thomas}{1999}]{Thomas}  Tikekar, R.,  Thomas, V. O. : {\it Pramana-J.  Phys.} {\bf 52} 237 (1999)
\bibitem[\protect\citeauthoryear{Patel \& Mehta}{1995}]{Patel} Patel, L. K.,  Mehta, N. P.: {\it Aust. J. Phys.} {\bf 48} 635 (1995)
\bibitem[\protect\citeauthoryear{Maharaj \& Maartens}{1989}]{Maartens}  Maharaj, S. D., Maartens, R.: {\it Gen. Rel. Grav.} {\bf 21} 899 (1989)
\bibitem[\protect\citeauthoryear{Gokhroo \& Mehra}{1994}]{Gokhroo} Gokhroo, M. K.,  Mehra, A. L. : {\it Gen. Rel. Grav.} {\bf  26} 75 (1994)
\bibitem[\protect\citeauthoryear{Mak \& Harko}{2003}]{Mak}  Mak, M. K., Harko, T.: {\it Proc. Roy. Soc. Lond. } {\bf A 459} 393 (2003)

\bibitem[\protect\citeauthoryear{Dev \& Gleiser}{2004}]{Dev}  Dev, K., Gleiser, M.: {\it Int. J. Mod. Phys.}{\bf D 13} 1389 (2004)
\bibitem[\protect\citeauthoryear{Chaisi \& Maharaj}{2005}]{Chaisi}  Chaisi, M.,  Maharaj, S. D. ; {\it Gen. Rel. Grav.} {\bf 37} 1177 (2005)
\bibitem[\protect\citeauthoryear{Bowers \& Liang}{1974}]{Bowers}  Bowers, R. L., Liang, E. P. T., {\it Astrophys. J.} {\bf  188} 657 (1974)
\bibitem[\protect\citeauthoryear{Sharma et al.}{2006}]{SK} Sharma, R.,  Karmakar, S.,  Mukherjee,S.: {\it Int. J. Mod. Phys. } {\bf  D 15} 405 (2006)
\bibitem[\protect\citeauthoryear{Karmakar et al.}{2007}]{SK1}  Karmakar, S., Sharma, R., Mukherjee, S.,  Maharaj, S. D.: {\it Pramana-J Phys.} {\bf 68} 881 (2007)
\bibitem[\protect\citeauthoryear{Mukherjee et al.}{1997}]{Mukherjee} Mukherjee,  S.,  Paul, B. C.,  Dadhich, N. K.: {\it Class. Quantum Grav.} {\bf 14} 3475 (1997)
\bibitem[\protect\citeauthoryear{Paul}{2004}]{bcpaul}  Paul, B. C.: {\it Int. J. Mod. Phys. } {\bf D 13} 229 (2004)
\bibitem[\protect\citeauthoryear{Paul}{2001}]{bcpaul1} Paul,  B. C.: {\it Class. Quantum Grav.} {\bf 18} 2637 (2001)
\bibitem[\protect\citeauthoryear{Vaidya \& Tikekar}{1982}]{Vaidya}  Vaidya, P. C.,  Tikekar, R.: {\it  J. Astrophys. Astron.} {\bf  3} 325 (1982)
\bibitem[\protect\citeauthoryear{Chattopadhyay et al.}{2012}]{b1} Chattopadhyay, P. K.,  Deb, R.,  Paul, B. C. : {\it Int. J. Mod. Phys.} {\bf D 21}   1250071 (2012)
\bibitem[\protect\citeauthoryear{Paul \& Deb}{2014}]{b2}  Paul B. C.,  Deb, R.:(communicated 2014)
\bibitem[\protect\citeauthoryear{Mafa Takisa \& Maharaj}{2013}]{maha}  Mafa Takisa, P.,   Maharaj,  S. D. : {\it  Gen. Rel. Grav. } {\bf 45} 1951 (2013)
\bibitem[\protect\citeauthoryear{Sharma \& Mukherjee}{2001}]{RS02}  Sharma, R., Mukherjee, S.: {\it  Mod. Phys. Lett.}{\bf A 16}  1049 (2001)
\bibitem[\protect\citeauthoryear{Sharma et al.}{2002}]{RS03}  Sharma, R.,  Mukherjee, S.,  Dey, M.,  Dey, J.: {\it  Mod. Phys. Lett.} {\bf A 17} 827 (2002)
\bibitem[\protect\citeauthoryear{Ponce de Leon}{2010}]{JP}  Ponce de Leon, J. : arXiv:1003.3151 [gr-qc]
\bibitem[\protect\citeauthoryear{Maharaj \& Mafa Takisa}{2012}]{Maharaj1}  Maharaj, S. D., Mafa Takisa, P.:  {\it Gen. Rel. Grav.} {\bf 44}, 1419 (2012)

\end{thebibliography}
\end{document}